\documentclass[aps,prd,twocolumn,superscriptaddress,a4paper]{revtex4}

\usepackage{graphicx}
\usepackage{eurosym}

\begin{document}


\title{Final Analysis and Results of the Phase II SIMPLE Dark Matter Search}


\author{M. Felizardo}
\affiliation{Department of Physics,
Universidade Nova de Lisboa, 2829-516 Caparica, Portugal}
\affiliation{Centro de F\'isica Nuclear, Universidade de Lisboa, 1649--003 Lisbon, Portugal}
\author{TA Girard}
\email[]{criodets@cii.fc.ul.pt} \affiliation{Centro de F\'isica
Nuclear, Universidade de Lisboa, 1649--003 Lisbon, Portugal}
\author{T. Morlat}
\affiliation{Ecole Normale Superieur de Montrouge, 1 Rue Aurice Arnoux,
92120 Montrouge, France}
\author{A.C. Fernandes} \affiliation{Instituto
Tecnol\'ogico e Nuclear, Instituto Superior T\'ecnico, Universidade T\'ecnica de Lisboa,
EN 10, 2686-953 Sacav\'em,
Portugal} \affiliation{Centro de F\'isica Nuclear, Universidade de
Lisboa, 1649--003 Lisbon, Portugal}
\author{A.R. Ramos} \affiliation{Instituto
Tecnol\'ogico e Nuclear, Instituto Superior T\'ecnico, Universidade T\'ecnica de Lisboa,
EN 10, 2686-953 Sacav\'em, Portugal} \affiliation{Centro de F\'isica Nuclear, Universidade de
Lisboa, 1649--003 Lisbon, Portugal}
\author{J.G. Marques}
\affiliation{Instituto Tecnol\'ogico e Nuclear, Instituto Superior T\'ecnico,
Universidade T\'ecnica de Lisboa, EN 10, 2686-953 Sacav\'em, Portugal} \affiliation{Centro de F\'isica
Nuclear, Universidade de Lisboa, 1649--003 Lisbon, Portugal}
\author{A. Kling}
\affiliation{Instituto Tecnol\'ogico e Nuclear, Instituto Superior T\'ecnico,
Universidade T\'ecnica de Lisboa, EN 10, 2686-953 Sacav\'em, Portugal} \affiliation{Centro de F\'isica
Nuclear, Universidade de Lisboa, 1649--003 Lisbon, Portugal}
\author{J. Puibasset }
\affiliation{CRMD-CNRS and Universit\'e d'Orl\'eans, 1 bis Rue de la Férollerie, 45071 Orl\'eans, France}
\author{M. Auguste}
\affiliation{Laboratoire Souterrain de Rustrel Pays d'Apt, UMS 3538 UNS/UAPV/CNRS, 84400 Rustrel, France}
\author{D. Boyer}
\affiliation{Laboratoire Souterrain de Rustrel Pays d'Apt, UMS 3538 UNS/UAPV/CNRS, 84400 Rustrel, France}
\author{A. Cavaillou}
\affiliation{Laboratoire Souterrain de Rustrel Pays d'Apt, UMS 3538 UNS/UAPV/CNRS, 84400 Rustrel, France}
\author{J. Poupeney}
\affiliation{Laboratoire Souterrain de Rustrel Pays d'Apt, UMS 3538 UNS/UAPV/CNRS, 84400 Rustrel, France}
\author{C. Sudre}
\affiliation{Laboratoire Souterrain de Rustrel Pays d'Apt, UMS 3538 UNS/UAPV/CNRS, 84400 Rustrel, France}
\author{H.S. Miley}
\affiliation{Pacific Northwest National Laboratory, Richland, WA
99352, USA}
\author{R.F. Payne}
\affiliation{Pacific Northwest National Laboratory, Richland, WA
99352, USA}
\author{F.P. Carvalho}
\affiliation{Instituto
Tecnol\'ogico e Nuclear, Instituto Superior T\'ecnico,
Universidade T\'ecnica de Lisboa, EN 10, 2686-953 Sacav\'em,
Portugal}
\author{M.I. Prud\^encio}
\affiliation{Instituto Tecnol\'ogico e Nuclear, Instituto Superior T\'ecnico,
Universidade T\'ecnica de Lisboa, EN 10, 2686-953 Sacav\'em,
Portugal}
\author{A. Gouveia}
\affiliation{Instituto Tecnol\'ogico e Nuclear, Instituto Superior T\'ecnico,
Universidade T\'ecnica de Lisboa, EN 10, 2686-953 Sacav\'em,
Portugal}
\author{R. Marques}
\affiliation{Instituto
Tecnol\'ogico e Nuclear, Instituto Superior T\'ecnico, Universidade T\'ecnica
de Lisboa, EN 10, 2686-953 Sacav\'em,
Portugal}

\collaboration{The SIMPLE collaboration} \noaffiliation

\date{\today}

\begin{abstract}
We report the final results of the Phase II SIMPLE measurements, comprising two run stages of
15 superheated droplet detectors each, the second stage including an improved neutron shielding.
The analyses includes a refined signal analysis, and revised nucleation efficiency based on reanalysis
of previously-reported monochromatic neutron irradiations. The combined results yield a contour
minimum of $\sigma_{p}$ = 5.7 $\times$ 10$^{-3}$ pb at 35 GeV/c$^{2}$ in the spin-dependent sector of WIMP-proton
interactions, the most restrictive to date for M$_{W} \leq$ 60 GeV/c$^{2}$ from a direct search experiment and overlapping for the
first time results previously obtained only indirectly. In the spin-independent sector, a minimum of
4.7 $\times$ 10$^{-6}$ pb at 35 GeV/c$^{2}$ is achieved, with the exclusion contour challenging a significant part
of the light mass WIMP region of current interest.
\end{abstract}


\maketitle


The search for weakly interacting massive particle (WIMP) dark matter remains at the forefront of modern physics activity. Estimated to comprise $\sim$ 23\% of the Universe mass, it is the role of direct detection efforts to elaborate its nature, and whether its interaction with nucleons is spin-independent (SI) or spin-dependent (SD). SIMPLE (Superheated Instrument for Massive ParticLe Experiments) \cite{prl} is a direct search activity using superheated liquid detectors, and one of only a few in the international panorama with sensitivity to the WIMP-proton sector of the SD phase space. It is operated at the 1500 mwe level of the Low Noise Underground Laboratory (LSBB) in southern France.

In \cite{prl}, we reported the first results of a two stage Phase II measurement, comprising a 14.1 kgd Stage 1 exposure of 15 superheated droplet detectors (SDDs) \cite{tomo,morlat,tmtese} with a total active mass of 0.208 kg. We here provide the results of the full Phase II measurement, including a 13.67 kgd Stage 2 exposure of a second 15 SDD set, together with improved neutron shielding and a refined analysis of the individual detector run signals, sensitivities, and nucleation efficiency.

A SDD consists of a dispersion of superheated liquid droplets homogeneously distributed within a gel matrix, which may undergo a transition to the gas phase upon energy deposition by incident radiation. Two conditions are required for the nucleation of the gas phase of the superheated droplets \cite{seitz}: (i) the energy deposited must be greater than a thermodynamic minimum, and (ii) this energy must be deposited within a thermodynamically-defined minimum distance ($\Lambda$r$_{c}$) inside the droplet, where $\Lambda$ is the nucleation parameter and r$_{c}$ = the thermodynamic critical bubble radius. Adjustment of the two conditions results in the necessity of depositions of order $\geq$ 150 keV/$\mu$m for a bubble nucleation, rendering the SDD effectively insensitive to the
majority of traditional detector backgrounds (including electrons, $\gamma$'s and cosmic muons) which complicate more conventional dark matter search detectors, leaving only $\alpha$- and neutron-induced events.

The 15 Stage 2 SDDs were fabricated as described in \cite{prl}, each containing between 11-19 g of C$_{2}$ClF$_{5}$ for a total
active mass of 0.215 kg; an additional, freon-less but otherwise identical, SDD again served as an acoustic veto. These were initially pressurized to 2.00$\pm$0.05 bar, and installed at the rate of one per day in a 700 liter water pool maintained at a bath temperature of 9.0$\pm$0.1 $^{\circ}$C, this time with the DAQ initiated only after the installation of each 8 detector set. The instrumentation was identical to that of Stage 1; in contrast to Stage 1 however, the SDD pressures were allowed to rise with time in order to obtain additional information on the measurement sensitivity.

Also in contrast to Stage 1, the water pool rested on an additional 10 cm of wood and paraffin, and 10 cm of polyethylene, with a rebuilt 50-75 cm thick surrounding water shield. As a result of the seasonal increase in water circulation within the mountain, the ambient radon level increased to $\sim$ 1000 Bq/m$^{3}$; continued purging of the cavern air reduced this to $\sim$ 100 Bq/m$^{3}$, and circulation of the pool water in combination with radioassays of the detector construction materials, yielded a Stage 2 $\alpha$-background estimate, including both progenitor and daughter decays, of 5.72 $\pm$ 0.12 (stat) $\pm$ 0.29 (syst) evt/kgd.

Extensive Monte-Carlo estimates of the expected neutron background, which accounted for spontaneous fission plus decay-induced ($\alpha$,n) reactions, and included the increased below-pool shielding and new materials radioassays, yielded a reduced rate of
0.333 $\pm$ 0.001 (stat) $\pm$ 0.038 (syst) evt/kgd; recalculation of the Stage 1 disposition with the new radioassays yielded a revised background rate of 0.976 $\pm$ 0.004 (stat) $\pm$ 0.042 (syst) evt/kgd, with the primary contribution being the concrete.
For the improved shielding of Stage 2, background neutrons originate mainly from the glass detector containment and shield water.

Stage 2 data was obtained between 12 April - 22 July 2010. The total exposure was 13.67 kgd, from the detector installation protocol and mechanical failure of 4 SDDs during the run as a result of overpressuring; no weather-induced data losses occurred.

Analysis of the Stage 2 signals, as per Stage 1, included a filtering of the initial data set (1997 events) via a pulse validation routine, a cross-correlation of the remaining set in time between all SDDs, and coincidence rejection as due to local noise events and that a WIMP interacts with no more than one of the in-bath detectors [1]; the analysis was improved via a new bandpass filter for noise suppression. The signal waveform, decay time constant and spectral density structure of the remaining 826 single events were next inspected individually. A particle-induced nucleation event possesses a characteristic frequency response, with a time span of a few milliseconds, a decay constant of 5-40 ms, and a primary harmonic between 0.45-0.75 kHz; these parameters differ significantly
from those of gel-associated acoustic backgrounds such as trapped N$_{2}$ gas, fractures and local acoustic backgrounds such as water bubbles \cite{felizardo}. The event-by-event analysis permits isolation of the particle-induced nucleation events with an efficiency of better than 97\% at 95\% C.L.

Figure 1 displays the signal amplitude and frequency for each of the identified 41 particle-induced signal events in Stage 2. Following calibrations as described in \cite{prl}, a nuclear recoil discrimination cut for A $\leq$ 100 mV was again imposed with an acceptance of $>$ 97\%, yielding a total of 2 events for the entire exposure.

The two bubble nucleation criteria are thermodynamic \cite{seitz}, so that variation of either temperature or pressure modifies
the recoil threshold energy and thus the SDD sensitivity, as seen in Fig. 2 where the expected variation in threshold recoil energies (E$_{thr}$) of both neutron-induced recoils and $\alpha$'s for several operating pressures is shown. The $\alpha$ threshold curve shifts to higher temperatures with increasing pressure. Since the curves depend on $\Lambda$ \cite{seitz}, comparison of experiment and theoretical predictions with varying $\Lambda$ confirmed our measurements \cite{jptese} of $\Lambda$ = 1.40 $\pm$0.05, yielding no $\alpha$-sensitivity whatsoever above 2.30 bar as observed experimentally. This was then used in calculating the ion recoil energy curves shown in Fig. 2. For pressures $\leq$ 2.20 $\pm$ 0.05 bar, the threshold recoil energy at 9$^{\circ}$C remains below 9.0 $\pm$ 0.3 keV.

\begin{figure}[h]
  \includegraphics[width=8 cm]{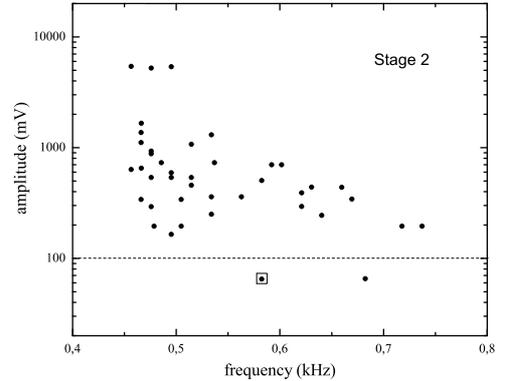}\\
  \caption{scatter plot of the amplitudes and frequency of the
primary harmonic of each true nucleation event observed over
the Stage 2 exposure, with the boxed event for pressures $\leq$ 2.2 bar.}
\label{fig3}
\end{figure}

\begin{figure}[h]
  \includegraphics[width=8 cm]{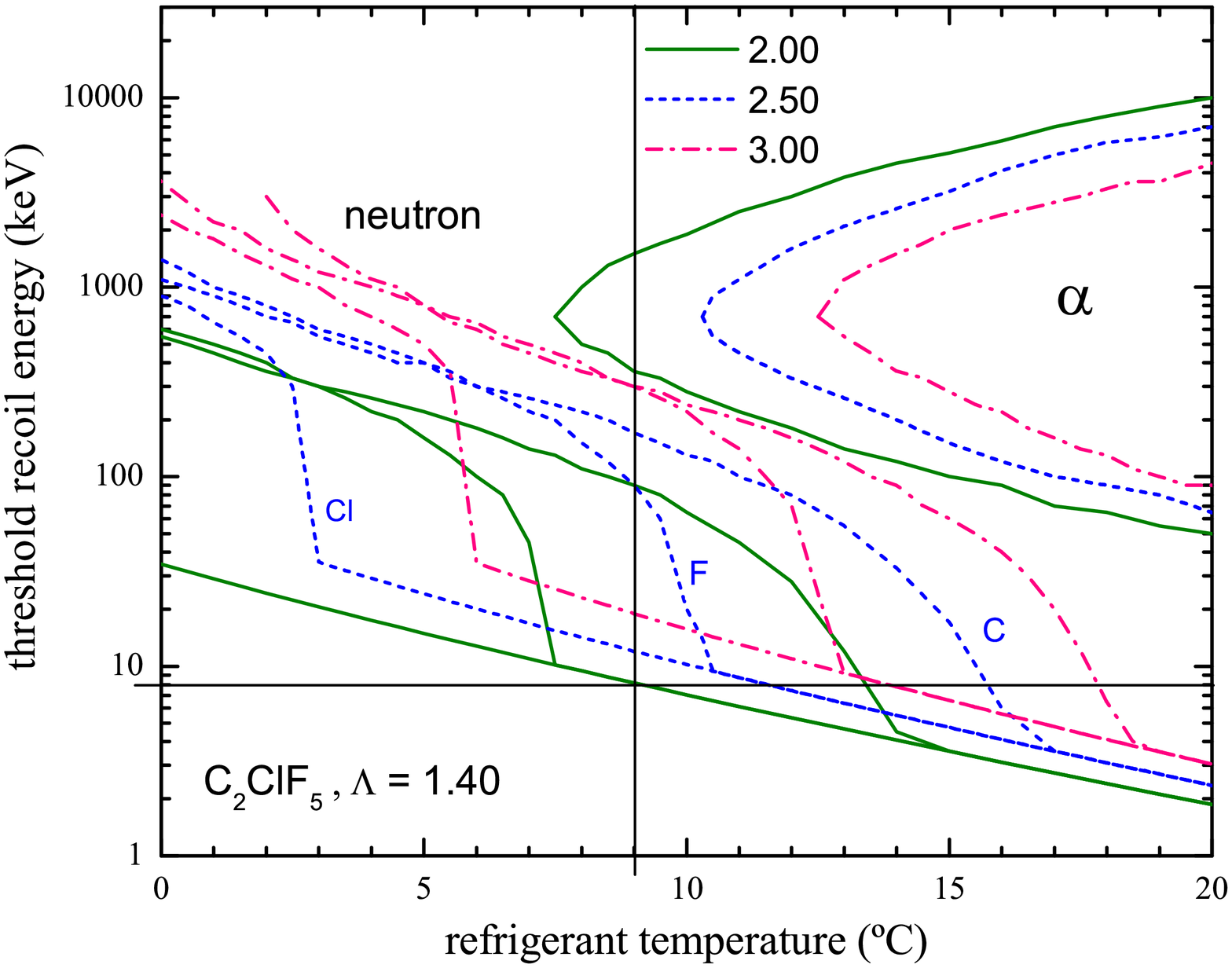}\\
  \caption{variation of the recoil and $\alpha$ energy thresholds
with temperature for the three C$_{2}$ClF$_{5}$ constituents at 2.00 (solid),
2.50 (dash) and 3.00 (dash-dot) bar, with $\Lambda$=1.40. The freon constituents
are identified for 2.5 bar. The vertical line indicates the 9$^{\circ}$C measurement
temperature; the horizontal line, a threshold recoil energy of 8 keV.}
\label{fig3}
\end{figure}

As also seen in Fig. 2, the ion recoil threshold curves similarly evolve to higher temperatures and energies with pressure increase, rendering the SDDs increasingly less responsive to the on-detector neutron spectrum: at 2.50 bar, this reduction is 30\%, consistent with the observed absence of any low amplitude events above 2.3 bar when weighted by the exposure.

The pressure records of all SDDs were next inspected for evolution during the measurement, and correlated with the signal records. Data obtained at pressures greater than 2.20 bar were excluded, reducing the Stage 2 exposure to 6.71 kgd; correlation with the signal record yielded 1 recoil event consistent with the estimated 2.2$\pm$0.3 background neutrons. The Stage 1 events were similarly pressure-correlated, reducing the exposure to 13.47 kgd; reanalysis of the recoil signals via a Hilbert transform-based demodulation identified 4 with exponential decay characteristic of nonuniform impulses observed in acoustic background studies associated with SDDs in vibrational contact with their support and air bubbles from water inflow, reducing the recoil events to 10, slightly below the estimated 13$\pm$0.6 background neutrons.

The first Stage 1 results resulted in part from a theoretical bubble nucleation efficiency given by $\eta$(E)=1-E$_{thr}$/E$_{dep}$ \cite{loapfel}. This $\eta$ however represents only a first approximation to the statistical nature of the energy deposition and its conversion into heat \cite{picasso2005}: a detailed reanalysis of previous monochromatic (54 and 149 keV) neutron irradiation data \cite{itn}, at 1 and 2 bar as a function of temperature, yielded a refined efficiency of $\eta$'=1-exp[-$\Gamma$(E/E$_{thr}$-1))] with $\Gamma$=4.2 $\pm$ 0.3, independent of pressure.

We show in Fig. 3 the impacts of the Stage 2 and reanalyzed Stage 1 results on SD WIMP-proton scattering, together with the competitive results of other direct \cite{picassotaup,coupptaup,kimstaup} and indirect \cite{desai,abbasi} experiments. The contours are calculated using the previous \cite{prl} Feldman-Cousins approach \cite{FC} based on observing n events against a background one systematic uncertainty below the estimated neutron-generated recoil background, $\eta$' with $\Gamma$=3.6, the standard isothermal halo and a WIMP scattering rate \cite{lewin} with zero momentum transfer, spin-dependent cross section $\sigma^{SD}_{p}$ for elastic scattering. The form factors of \cite{lewin} have been used for all odd-A nuclei, with the spin values of \cite{pacheco} used for $^{19}$F; for $^{35}$Cl and $^{37}$Cl, the spin values are from \cite{mi}, while for $^{13}$C they were estimated using the odd group approximation. The Stage 2 result is seen to nearly equal the revised Stage 1 result with its revised minimum of $\sigma_{p}$ = 9.2 $\times$ 10$^{-3}$ pb at 35 GeV/c$^{2}$, despite half the exposure.

The above representation neglects the non-negligible spin contribution of the neutron sector in $^{19}$F, which is captured in a model-independent SD formulation \cite{mi} with $\sigma_{SD}$ $\sim$ [a$_{p}<$S$_{p}>$+a$_{n}<$S$_{n}>$]$^{2}$, where a$_{p,n}$ are the WIMP-proton,neutron coupling strengths, and $<$S$_{p,n}>$ are the expectation values of the proton (neutron) group spins. In this representation, experiments define a band (single nuclei targets) or an ellipse (multi-nuclei target), with the allowed area defined by the intersect of the most sensitive results in a$_{p}$, a$_{n}$. At M$_{W}$ = 50 GeV/c$^{2}$, combined with neutron-sensitive XENON10 \cite{xenon2008}, the allowed area reduction is better than 2/3 compared with Ref. \cite{prl}; masses above or below this choice yield slightly increased limits for most all experiments. More relevant would however be the model-independent results for M$_{W}$ $\sim$ 10 GeV/c$^{2}$, unavailable for the majority of experiments.

\begin{figure}[h]
  \includegraphics[width=9 cm]{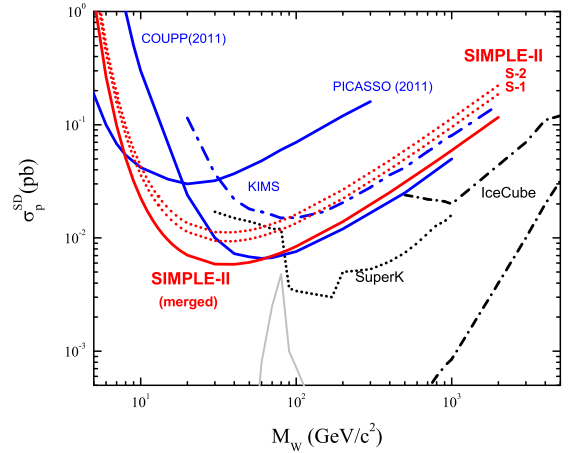}\\
  \caption{various spin-dependent WIMP-proton exclusion contours
for Phase II, together with the leading direct \cite{picassotaup,coupptaup,kimstaup}
and indirect SuperK \cite{desai}, IceCube \cite{abbasi} search results; shown
are the Stage 2 result, the reanalyzed Stage 1 result,  and a merging of the
two. The region outlined in grey is favored by cMSSM \cite{roszkowski}.} \label{fig4}
\end{figure}

The impact of the results in the SI sector is shown in Fig. 4 in comparison with results from other leading search
efforts \cite{coupptaup,kimstaup,xenon2008,xenon2011,xenon2011a,cdms2010,cdms2011,cresst2,zeplin2009,zeplin2010,dama,picasso2012,cogent2011}, again calculated with the standard isothermal halo and WIMP elastic scattering rate of Ref. \cite{lewin} using Feldman-Cousins, a Helm nuclear form factor, and $\eta$'. Again, the Stage 2 contour is nearly equal to the revised Stage 1 contour with its contour minimum of 7.6$\times$10$^{-6}$ pb at 35 GeV/c$^{2}$. Owing to the low recoil energy threshold, both results enter the possible light mass WIMP region recently suggested by CoGeNT \cite{cogent2011} and CRESST-II \cite{cresst2}.

A straightforward combination of the two results using the Feldman-Cousins approach, based on 11 candidates with an assumed background 1 $\sigma$(syst) below the expected total background, yields the "merged" contours indicated in each of Figs. 3 and 4; in the SI case, the contour minimum drops to 4.7$\times$10$^{-6}$ pb and the result is in tension with the recent reports of CoGeNT \cite{cogent2011}, DAMA/LIBRA \cite{dama} and CRESST \cite{cresst2} regarding light mass WIMPS, using a significantly different technique with different systematics than the XENON \cite{arxiv1} and CDMS \cite{arxiv2} experiments. For the case of SD interactions, the contour minimum drops to 5.7$\times$10$^{-3}$ pb, constituting the most restrictive direct search limit on SD WIMP-proton scattering for M$_{W} \leq$ 60 GeV/c$^{2}$ to date, and beginning to complement the more sensitive results obtained by indirect detection measurements.

\begin{figure}[h]
  \includegraphics[width=10 cm]{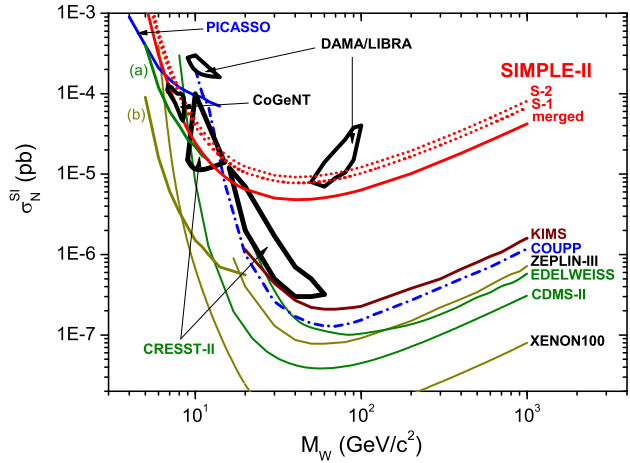}\\
  \caption{various spin-independent contours for Phase II, together with those of the leading \cite{xenon2008,xenon2011,xenon2011a,cdms2010,cdms2011,coupptaup,kimstaup, picasso2012,zeplin2009,zeplin2010} spin-independent
search results; shown are the Stage 2 result, the reanalyzed Stage 1 result, and a merging of the two.
The partial contours (a) and (b) are taken from \cite{cdms2011} and \cite{xenon2011}, respectively. The closed areas identified as either CRESST-II \cite{cresst2}, DAMA/LIBRA \cite{dama} or CoGeNT \cite{cogent2011} represent the regions in which possible light mass WIMPS have been respectively reported.}
  \label{fig5}
\end{figure}

The improved restrictions of the revised Stage 1 contour are a direct result of the more detailed signal analysis, improved radio-assays of the shielding materials, and the revised nucleation efficiency in the analysis: Stage 2, with the additional benefit of its improved neutron shielding, provides an almost identical sensitivity with half the Stage 1 exposure. While the merging may be questioned, the results are sufficient motivation for a larger exposure measurement with further neutron background reduction, and variation of the SDD operating temperature/pressure to provide a lower recoil energy threshold, towards clarifying the situation. Variation of the refrigerant between C$_{3}$F$_{8}$, C$_{4}$F$_{8}$, CF$_{3}$I or one of the other SDDs developed by SIMPLE in recent years \cite{tomo} allows a variation of detector sensitivities between SI and SD sectors \cite{bertone}. An energy spectrum can in principle be obtained, should candidate events be identified, by either a temperature or pressure ramping of the SDDs.

The suggested improvements in this measurement however also require a significantly increased active target mass in order to be competitive, for which the fabrication requirements of the current SDDs become unmanageable: a new device, originally prototyped in 2000 \cite{jptese} and re-prototyped in 2010, based on a large superheated freon droplet contained within a gel-sheathed vessel (effectively a bubble chamber), is currently completing development; it will permit a factor 25 increase in the active mass with reduced space requirements. An additional all-around 60 cm of purified water shielding increases the neutron suppression by more than 10$^{3}$, giving the possibility to achieve exposures of 10$^{2}$ kgd in a few weeks.

We again thank Dr. F. Giuliani (U. New Mexico) for numerous suggestions and advices, Dr. P. Loaiza (LSM) for her several radioassays of the site concrete and steel, as well as Eng. J. Albuquerque (CRIOLAB, Lda) for his various construction assistances during the work staging, and Mr. A. da Silva (ACP) for his many transport interventions. We also thank the 12 students of Dr. S. Gaffet's 2010 geophysics class for their assistance in the Stage 2 shielding reconstruction, and the Casoli's, Escoffier's, L'Aptois and Mairie of Rustrel for their kind hospitalities and support during our various residences near the LSBB. This work was supported by the Nuclear Physics Center of the University of Lisbon, and by grants PDTC/FIS/83424/2006 and PTDC/FIS/115733/2009 of the Portuguese Foundation for Science and Technology (FCT). M. Felizardo was supported by SFRH/BD/46545/2008 of FCT.



\begin{thebibliography}{21} \expandafter\ifx\csname
natexlab\endcsname\relax\def\natexlab#1{#1}\fi
\expandafter\ifx\csname bibnamefont\endcsname\relax
  \def\bibnamefont#1{#1}\fi
\expandafter\ifx\csname bibfnamefont\endcsname\relax
  \def\bibfnamefont#1{#1}\fi
\expandafter\ifx\csname citenamefont\endcsname\relax
  \def\citenamefont#1{#1}\fi
\expandafter\ifx\csname url\endcsname\relax
  \def\url#1{\texttt{#1}}\fi
\expandafter\ifx\csname urlprefix\endcsname\relax\def\urlprefix{URL
}\fi \providecommand{\bibinfo}[2]{#2}
\providecommand{\eprint}[2][]{\url{#2}}

\bibitem[{\citenamefont{Felizardo}(2010)}]{prl}
\bibinfo{author}{\bibfnamefont{M.} \bibnamefont{Felizardo}} \bibnamefont{et.~al.},
\bibinfo{journal}{Phys. Rev. Lett.} \bibinfo{volume}{105},
\bibinfo{pages}{211301} (\bibinfo{year}{2010}).

\bibitem[{\citenamefont{Morlat}(2006)}]{tomo}
\bibinfo{author}{\bibfnamefont{T.}~\bibnamefont{Morlat}}
\bibnamefont{et.~al.}:
\bibinfo{journal}{Nucl. Instr. \& Meth.} \bibinfo{volume}{A560},
\bibinfo{pages}{339} (\bibinfo{year}{2006}).

\bibitem[{\citenamefont{Morlat}(2008)}]{morlat}
\bibinfo{author}{\bibfnamefont{T.}~\bibnamefont{Morlat}} \bibnamefont{et.~al.}
\bibinfo{journal}{Astrop. Phys.} \bibinfo{volume}{30},
\bibinfo{pages}{159}(\bibinfo{year}{2008}).

\bibitem[{\citenamefont{Morlat}(2000)}]{tmtese}
\bibinfo{author}{\bibfnamefont{T.} \bibnamefont{Morlat}}:
\bibinfo{journal}{PhD thesis, Univ. Paris 7, 2004 (unpublished)}.

\bibitem[{\citenamefont{Seitz}(1958)}]{seitz}
\bibinfo{author}{\bibfnamefont{F.} \bibnamefont{Seitz}},
\bibinfo{journal}{Phys. Fluids} \bibinfo{volume}{1},
\bibinfo{pages}{1} (\bibinfo{year}{1958}).

\bibitem[{\citenamefont{Felizardo}(2009)}]{felizardo}
\bibinfo{author}{\bibfnamefont{M.} \bibnamefont{Felizardo}}
\bibnamefont{et.~al.},
\bibinfo{journal}{Nucl. Instrum. Meth.} \bibinfo{volume}{A589},
\bibinfo{pages}{72} (\bibinfo{year}{2008}).

\bibitem[{\citenamefont{Puibasset}(2000)}]{jptese}
\bibinfo{author}{\bibfnamefont{J.} \bibnamefont{Puibasset}}:
\bibinfo{journal}{PhD thesis, Univ. Paris 7, 2000 (unpublished)}.

\bibitem[{\citenamefont{loapfel}(1996)}]{loapfel}
\bibinfo{author}{\bibfnamefont{Y. C.}~\bibnamefont{Lo}},
\bibinfo{author}{\bibfnamefont{R. E.}~\bibnamefont{Apfel}}:
\bibinfo{journal}{Phys. Rev.} \bibinfo{volume}{A38},
\bibinfo{pages}{5260} (\bibinfo{year}{1988}).

\bibitem[{\citenamefont{M. Barnab\'e-Heider}(2005)}]{picasso2005}
\bibinfo{author}{\bibfnamefont{M.}~\bibnamefont{Barnab\'e-Heider}} \bibnamefont{et. al.},
\bibinfo{journal}{Nucl. Instr. \& Meth.} \bibinfo{volume}{555},
\bibinfo{pages}{184} (\bibinfo{year}{2005}).

\bibitem[{\citenamefont{Giuliani}(2004)}]{itn}
\bibinfo{author}{\bibfnamefont{F.} \bibnamefont{Giuliani}} \bibnamefont{et.~al.},
\bibinfo{journal}{Nucl. Instr. Meth.} \bibinfo{volume}{A526},
\bibinfo{pages}{526} (\bibinfo{year}{2004}).

\bibitem[{\citenamefont{Zacek}(2011)}]{picassotaup}
\bibinfo{author}{\bibfnamefont{V.} \bibnamefont{Zacek}}:
\bibinfo{journal}{talk at TAUP-2011 Workshop, Munich, Germany, Sept. 5-9, 2011}.

\bibitem[{\citenamefont{Lippincott}(2011)}]{coupptaup}
\bibinfo{author}{\bibfnamefont{W.H.} \bibnamefont{Lippincott}}:
\bibinfo{journal}{talk at TAUP-2011 Workshop, Munich, Germany, Sept. 5-9, 2011}.

\bibitem[{\citenamefont{Kim}(2011)}]{kimstaup}
\bibinfo{author}{\bibfnamefont{S.K.} \bibnamefont{Kim}}:
\bibinfo{journal}{talk at TAUP-2011 Workshop, Munich, Germany, Sept. 5-9, 2011}.

\bibitem[{\citenamefont{Desai}(2004)}]{desai}
\bibinfo{author}{\bibfnamefont{S.}~\bibnamefont{Desai}} \bibnamefont{et.~al.},
\bibinfo{journal}{Phys. Rev.} \bibinfo{volume}{D70},
\bibinfo{pages}{083523} (\bibinfo{year}{2004}).

\bibitem[{\citenamefont{Abbasi}(2009)}]{abbasi}
\bibinfo{author}{\bibfnamefont{R.}~\bibnamefont{Abbasi}} \bibnamefont{et.~al.},
\bibinfo{journal}{Phys. Rev. Lett.} \bibinfo{volume}{102},
\bibinfo{pages}{201302} (\bibinfo{year}{2009}).

\bibitem[{\citenamefont{Fel-}(1996)}]{FC}
\bibinfo{author}{\bibfnamefont{G. J.}~\bibnamefont{Feldman}},
\bibinfo{author}{\bibfnamefont{R. D.}~\bibnamefont{Cousins}}:
\bibinfo{journal}{Phys. Rev.} \bibinfo{volume}{D57},
\bibinfo{pages}{3873} (\bibinfo{year}{1998}).

\bibitem[{\citenamefont{roszkowski}(2007)}]{roszkowski}
\bibinfo{author}{\bibfnamefont{L.}~\bibnamefont{Roszkowski}} \bibnamefont{et.~al.},
\bibinfo{journal}{J. High Energy Phys} \bibinfo{volume}{07},
\bibinfo{pages}{075} (\bibinfo{year}{2007}).

\bibitem[{\citenamefont{Lewin}(1996)}]{lewin}
\bibinfo{author}{\bibfnamefont{J.~D.}~\bibnamefont{Lewin}} \bibnamefont{and}
\bibinfo{author}{\bibfnamefont{P.~F.} \bibnamefont{Smith}},
\bibinfo{journal}{Astrop. Phys.}  \bibinfo{volume}{6},
\bibinfo{pages}{87}(\bibinfo{year}{1996}).

\bibitem[{\citenamefont{pacheco}(1989)}]{pacheco}
\bibinfo{author}{\bibfnamefont{A.~F.}~\bibnamefont{Pacheco}} \bibnamefont{and}
\bibinfo{author}{\bibfnamefont{D.} \bibnamefont{Strottman}},
\bibinfo{journal}{Phys. Rev.} \bibinfo{volume}{D40},
\bibinfo{pages}{2131}(\bibinfo{year}{1989}).

\bibitem[{\citenamefont{Giuliani}(2005)}]{mi}
\bibinfo{author}{\bibfnamefont{F.}~\bibnamefont{Giuliani}} \bibnamefont{and}
\bibinfo{author}{\bibfnamefont{TA} \bibnamefont{Girard}},
\bibinfo{journal}{Phys. Rev.} \bibinfo{volume}{D71},
\bibinfo{pages}{123503}(\bibinfo{year}{2005}).

\bibitem[{\citenamefont{Angle}(2008)}]{xenon2008}
\bibinfo{author}{\bibfnamefont{J.}~\bibnamefont{Angle}} \bibnamefont{et.~al.},
\bibinfo{journal}{Phys. Rev. Lett.} \bibinfo{volume}{100},
\bibinfo{pages}{021303} (\bibinfo{year}{2008}).

\bibitem[{\citenamefont{Angle}(2011)}]{xenon2011}
\bibinfo{author}{\bibfnamefont{J.}~\bibnamefont{Angle}}
\bibnamefont{et.~al.},
\bibinfo{journal}{arxiv:1104.3088v1 [astro-ph.CO]}.

\bibitem[{\citenamefont{Aprile}(2011)}]{xenon2011a}
\bibinfo{author}{\bibfnamefont{E.} \bibnamefont{Aprile}}
\bibnamefont{et.~al.},
\bibinfo{journal}{arxiv:1104.2549v1 [astro-ph.CO]}.

\bibitem[{\citenamefont{Ahmed}(2010)}]{cdms2010}
\bibinfo{author}{\bibfnamefont{Z.}~\bibnamefont{Ahmed}}
\bibnamefont{et.~al.},
\bibinfo{journal}{Science} \bibinfo{volume}{327},
\bibinfo{pages}{1619} (\bibinfo{year}{2010}).

\bibitem[{\citenamefont{Ahmed}(2011)}]{cdms2011}
\bibinfo{author}{\bibfnamefont{Z.}~\bibnamefont{Ahmed}}
\bibnamefont{et.~al.},
\bibinfo{journal}{Phys. Rev. Lett.} \bibinfo{volume}{106},
\bibinfo{pages}{131302} (\bibinfo{year}{2011}).

\bibitem[{\citenamefont{Lebedenko}(2001)}]{zeplin2009}
\bibinfo{author}{\bibfnamefont{V.N.}~\bibnamefont{Lebedenko}}
\bibnamefont{et.~al.},
\bibinfo{journal}{Phys. Rev.} \bibinfo{volume}{D80},
\bibinfo{pages}{052010} (\bibinfo{year}{2009}).

\bibitem[{\citenamefont{Armengaud}(2010)}]{zeplin2010}
\bibinfo{author}{\bibfnamefont{E.}~\bibnamefont{Armengaud}}
\bibnamefont{et.~al.},
\bibinfo{journal}{Phys. Lett.} \bibinfo{volume}{B687},
\bibinfo{pages}{294} (\bibinfo{year}{2010}).

\bibitem[{\citenamefont{Archambault}(2011)}]{picasso2012}
\bibinfo{author}{\bibfnamefont{S.} \bibnamefont{Archambault}}
\bibnamefont{et.~al.},\bibinfo{journal}{ arXiv:1202.1240v1 [hep-ex].}

\bibitem[{\citenamefont{Angolher}(2011)}]{cresst2}
\bibinfo{author}{\bibfnamefont{G.}~\bibnamefont{Angloher}} \bibnamefont{et.~al.},
\bibinfo{journal}{arXiv:1109.0702v1 [astro-ph.CO].}

\bibitem[{\citenamefont{Savage}(2009)}]{dama}
\bibinfo{author}{\bibfnamefont{C.}~\bibnamefont{Savage}} \bibnamefont{et.~al.},
\bibinfo{journal}{JCAP}
\bibinfo{volume}{0904},
\bibinfo{pages}{010} (\bibinfo{year}{2009}).

\bibitem[{\citenamefont{Aalseth}(2011)}]{cogent2011}
\bibinfo{author}{\bibfnamefont{C.E.}~\bibnamefont{Aalseth}}
\bibnamefont{et.~al.},
\bibinfo{journal}{Phys. Rev. Lett.} \bibinfo{volume}{106},
\bibinfo{pages}{131301} (\bibinfo{year}{2011}).

\bibitem[{\citenamefont{Collar}(2011)}]{arxiv1}
\bibinfo{author}{\bibfnamefont{J.I.} \bibnamefont{Collar}}:
\bibinfo{journal}{arxiv:1106.0653v1 [astro-ph.CO].}
(\bibinfo{year}{2011}).

\bibitem[{\citenamefont{Collar}(2011)}]{arxiv2}
\bibinfo{author}{\bibfnamefont{J.I.} \bibnamefont{Collar}}:
\bibinfo{journal}{arxiv:1103.3481v1 [astro-ph.CO].}
(\bibinfo{year}{2011}).

\bibitem[{\citenamefont{Bertone}(2007)}]{bertone}
\bibinfo{author}{\bibfnamefont{G.}~\bibnamefont{Bertone}},
\bibinfo{journal}{Phys. Rev. Lett.} \bibinfo{volume}{99},
\bibinfo{pages}{151301} (\bibinfo{year}{2007}).

\end{thebibliography}

\end{document}